\documentstyle[12pt]{article}

\begin{document}
\topmargin -15pt
\oddsidemargin 5mm
\setcounter{page}{1}
\vspace{2cm}
\begin{center}
{\bf Do The Particles With Nonzero Mass Need Duality ?}\\
\vspace{3mm}
{\large  Genrikh Bayatian}\\
\vspace{3mm}
Yerevan Physics Institute\\
Alikhanian Brothers St. 2, Yerevan 375036 Armenia\\

\end{center}

\vspace{3mm}
\begin{abstract}
Unsufficiency of conditions for the formation of interference in
monoparticle "which path" experiments is proven. The version of 
corpuscular interpretation of diffraction pattern based on the action 
discreteness is presented, which makes unnecessary to accredit duality 
to the particles with nonzero mass. The experiment to check the above 
mentioned concepts are proposed.
\end{abstract}

{\bf 1. Introduction.}\\
\indent
Interpreting all the "which path" experimental data one usually
 proceeds from the fact that the observed irregularities at single 
 microparticle diffraction have interferencial origin in spite of the 
 lack of strictly definitive experimental evidence. The resemblance of 
 diffraction pattern in multiphoton (classical optics) and monophoton 
 experiments may appear a scanty proof of interference presence in the 
 latter case.
 
 \indent
It is an axiom that simultaneous arrival of at least two monochromatic
and coherent waves to the observation point is an obligatory condition
for the formation of interference.
Observing no second wave in monophoton (monoparticle) experiments and 
basing on the generally accepted understanding of the interferencial
origin of mentioned irregularities most of the researchers were 
forced to assume the interference of a single particle (photon) with 
itself by its violent simultaneous pulling through both slits 
separated with spatial distance by several orders longer (by the order
of 4 for the experiments with  atoms of helium [1]) than the transverse 
sizes of these particles. Meanwhile, the particles do not 
undergo fission: no loss of microparticle energy or photon frequency 
is observed behind the slits.

\indent
This paper shows that the registered irregularities in monoparticle 
(monophoton) "which path" experiments may not have interferencial origin due
 to the insufficicency conditions for the interference realization 
 (two waves as minimum). This is why the attempts to interpret diffraction 
 pattern in monophoton (monoparticle) experiments using interference 
 (maybe only seeming) bring to well-known logical contradictions or, 
 according Feynman, to a puzzle [2].
 
\indent
In order to discover a logical hitch and to provide obligatory conditions 
for the interference we are forced using superposition principle and 
diffraction grating  to fission a single photon into two ones [3], 
ignoring the law of conservation of energy and other quantum numbers 
(lepton and hadron numbers in case of diffraction of electrons,
 nucleons and nuclei).
 
\indent
The only one way out of this logical deadlock is may seem to withdraw 
seeming interference because of scanty conditions for its realization 
in the monoparticle "which path" experiments and to search for new 
mechanisms of origin of detected irregularities from the corpuscular 
point of view by analogy to photo- and Compton-effects, when one has to
use corpuscular representations, although nobody doubts in the wave 
properties of photons.

{\bf 2.  The version of corpuscular interpretation.}\\
\indent
In our recent publication [3] a brief description of this interpretation
is given. Due to the importance of the problem we are giving below more
 detailed presentation.
 
 \indent
In the diffraction experiments the corpuscular-field interactions of 
microparticles with matter of grating are not taken into account. 
Only the wave-diffractional aspects are regarded. Such approach is
 acceptable in the classical optics, where out of a huge flux of photons 
 simultaneously dropping on the diffraction grating, two or more photons 
 will appear, which passed through the different slits. Meeting each other 
 they will interfere here among themselves. Although in this case 
 the mechanism of light beam deflection from the 
 initial direction is 
 not clear enough without accounting of Huygens-Fresnel approximate model,
  which should be corrected in the light of corpuscular presentation.
Quite different situation is observed, when a single photon, which may 
interfere only with itself (a puzzle!) [2] falls on two slits with 
$ b\gg\lambda$ 
($b$ is the distance between the slits and $\lambda$ is the 
length of photon wave).

\indent
In order to understand why the corpuscular-field interpretation is the only 
correct way out to solve the puzzle we shall present the experimental results. 
The possibility of diffractional scattering of electrons and neutrons with the 
identical de Broglie wave lengths at one and the same crystal differs by 
the order of  $\sim 7$ to the electrons. The electron interacts with the crystal 
through Coulomb field, while the neutron - mainly through the short-range
nuclear forces.
Such substantial difference at the same $\lambda /b$ proportions shows
 the dominant 
 importance of corpuscular-field interactions as compared with the 
 wave-diffractional ones.
 
 \indent
The account of microparticle interaction with the grating matter becomes 
more simple, if we accept the action multiplicity from the Planck constant 
($h$) for the non-bound states as well. This hypothesis has no heuristical value.
It is used for a long time in non-explicit form in theoretical analysis
of microparticle scattering angular distribution using the method of
partial waves [4] in the form of discreteness angular momentum having 
dimension of action.

\indent
The discrete behaviour of the action is natural because of the fact,
 that the non-bound state may differ in principle from its analogue 
 in bound state in as less "surplus" as the kinetic energy. The nature 
 is hardly arranged in the manner, when, e.g. at Coulomb interaction of 
 electron with proton in hydrogen atoms the action is quantized, while at 
 Coulomb scattering of the same particles one over another near to kinematic
  region of  hydrogen atoms the action will not be quantized.
  
  \indent
Let a parallel beam of microparticles with $P$ momentum   falls on the 
crystal with $b$ period and as a result of elastic scattering obtains $P_{r}$ 
transverse momentum. The scattering angle $\theta$ is 
determined by the relation
\begin{equation}
\label{AA}
P\sin \theta=P_{r}~.
\end{equation}
It is necessary to find out $P_{r}$ from the following differential equation 
\begin{equation}
\label{AB}
d\vec{P_{r}}=\vec{F_{r}}dt
\end{equation}
where $F_{r}$ is the force acting on the particle from the crystal in the 
direction perpendicular to the beam. Scalaraly multiplying [2] by $d\vec{r}$ of
path in $\vec{F_{r}}$ direction we shall receive 
\begin{equation}
\label{AG}
d\vec{P_{r}}\cdot d\vec{r}=\vec{F_{r}}\cdot d\vec{r}\cdot dt=dS_{r}(r,t)
\end{equation}
where $dS_{r}$ - is the action in $\vec{F_{r}}$
direction of path $d\vec{r}$ in $dt$ time period.

\indent
In order to receive complete action during $t$ time when the particle is in
crystal it is necessary to define the integration limits of $r$ and $t$ variables. 
Choosing the limits for $r$ from 0 to $b$ and for the time period from 0 to 
$t$ the equation (3) may be rewritten as 
\begin{equation}
\label{AD}
\int_{0}^{t}dP_{r} \int_{0}^{b}dr=\int_{0}^{t} \int_{0}^{b}dS_{r}(r,t)
\end{equation}
With the account that the right side is the complete action and,
as we suppose, should be equal to the multiple of Planck constant $h$, 
we may have 
\begin{equation}
\label{AF}
P_{r}\cdot b=nh
\end{equation}
If we substitute from here the value $P_{r}$ in (1), 
we shall receive 
\begin{equation}
\label{AJ}
b\cdot\sin\theta=n\cdot h/P=n\lambda
\end{equation}
 where $\lambda=h/P$.
The formula (6) shows, that the quantitative agreement with de Broglie
 hypothesis observed in the experiment is not accidental. 
 The de Broglie hypothesis contains in a latent form our hypothesis [3] 
 about the action discreteness of $h$ in all types of interactions.
  Due to the latter fact, as it may be seen from (6), the diffraction 
scattering angles $\theta$ obtain discrete values, imitating interferencial pattern.
The corpuscular-field interpretation, which leads to the same quantitative 
results as the de Broglie wave diffractional representations, allows to 
avoid generally known logical difficulties [2] at the interpretation 
of the diffractional pattern. This removes the discussed contradictions 
[3] of the superposition principle to the assumption of interference
 in the monoparticle "which path" experiments. This principle is 
 apparently is not acceptable in the case of alternative and mutually 
 transitional in time states, such as the transition of a single particle 
 either through the first or the second slit. Such events are unable to
  interfere due to the absence of partner. The composition of their 
  probability amplitudes is impossible, because they will compose the
   mixture combination of the probabilities. 

{\bf 3.  Expected experimental evidences.}\\
\indent
3.1.  The analysis of existing experiments.\\
\indent
In [3] we have proposed the real version of two slit Young type experiment
 [1], capable to give the conclusive description of the effect of open slit, 
 that has not passed by microparticle, on its behaviour. It is proposed
 to compare the diffraction pattern of two open slits with a total picture,
 when either one or another slit opens in its turn. In the total exposure 
 the interference must not appear by definition.
 
\indent
It is easy to propose that under $b\gg\lambda$ conditions 
the pattern with two open 
slits will be identical to the total picture, as far as the single 
microparticle will pass through one of two slits, while the other slit,
 which is not passed, may be at that moment considered as closed. Hence, 
 the interference should be not possible at two slits either in the 
 experiments with single particles. While the irregulations observed in
[1, 5, 6] may be interpreted by the formula (6). However, this does not 
mean that the experiment should not be done. It will finally prove what
 is correct.
 
 \indent
 3.2.  New possible experiments.
 
 \indent
 In order to prove the impossibility of a single microparticle transit 
through two slits (and therefore the impossibility of interference) we 
propose the experiment based on the Young scheme (or its analogy). 
Let us dispose detectors one by one after each slit and direct their
 signals to the coincidence scheme. The absence of coincidence signals 
 will evidence of the impossibility of simultaneous passing of a single 
 particle through the both slits, while the existence of such signals 
 will prove the opposite concept.
 
 \indent
 Let us discuss the scheme of one more experiment. The parallel beam of 
coherent photons using semitransparent mirror is fissioned into two parts: 
reflected and transmitted ones. Arranging the meeting of these two parts
 we shall see the interferencial picture.
 
 \indent
 Then using the optical filter we create out of multiphoton beam a 
single photon beam with $t_{m}\ll t_{c}$ 
parameters, where $t_{m}$ is a mean time interval between 
neighbour photons and $t_{c}$ is the train duration. In this case there 
will be no interference, because the photons will enter the observation 
zone at different times.

\indent
This one and two slit Young type experiments have the same origin. 
In both cases the photon has two possible unpredictable paths. The 
advantage of such version, based on the absence of the interference 
imitation conditioned on formula (6), will serve as a touchstone for
 the corpuscular interpretation  [3] of diffraction pattern, in contrast 
 to the previous tests, when only the legitimacy (correctness) of the wave 
 representations (interpretations) could be resolved.

{\bf Conclusion.}\\
\indent
The causality principle excludes the emergence of interference in the
 single particle (monophoton) "which path" experiments. Proceeding from 
 this it is impossible to interpret the irregularities observed in these 
 experiments using wave representations. They are likely to be connected 
 with the discreteness of action and for non-bound states.
 
\indent
The experiment according the described schemes are necessary in order
to prove the correctness of the proposed fundamental statement together
 with corpuscular interpretation.

{\bf Acknowledgments.}\\
It is a pleasure to thank Professor E.Mamidjanian for useful discussions.

\end{document}